\documentclass[amstex,epsf,amssymb,usenatbib]{mn2e}
\usepackage{epsf}
\usepackage{epsfig}
\usepackage{amsmath,amssymb}

\def\araa{ARA\&A}             
\def\apj{ApJ}                 
\def\apjl{ApJ}                
\def\apss{Ap\&SS}             
\def\aap{A\&A}                
\def\mnras{MNRAS}             
\newcommand{\Msolar}{\mbox{\,$\rm M_{\odot}$}}        
\newcommand{\Rsolar}{\mbox{\,$\rm R_{\odot}$}}        

\title{On the angular momentum evolution of merged white dwarfs}

\author[K.N. Gourgouliatos \& C.S. Jeffery]
       {K.N. Gourgouliatos\thanks{Present address: Institute of
       Astronomy, The Observatories, Madingley Road, Cambridge CB3 0HA}\thanks{E-mail: kng22@cam.ac.uk}
\& C.S. Jeffery\thanks{E-mail: csj@arm.ac.uk} \\
Armagh Observatory, College Hill, Armagh BT61 9DG, Northern Ireland
}

\date{Accepted .....
      Received ..... ;
      in original form .....}

\pagerange{\pageref{firstpage}--\pageref{lastpage}}
\pubyear{2006}

\begin{document}

\maketitle

\label{firstpage}

\begin{abstract}
\noindent
We study the angular momentum evolution of binaries containing two
white dwarfs which merge and become cool helium-rich supergiants. 
Our object is to compare predicted rotation velocities with
observations of highly evolved stars believed to have formed from such
a merger, which include the R\,CrB and extreme helium stars. 

The principal case study involves a short-period binary containing a 
0.6 $M_{\sun}$ carbon-oxygen white dwarf, and a 0.3 $M_{\sun}$  helium
white dwarf. The initial condition for the angular momentum
distribution is defined by the orbital
configuration where the secondary fills its Roche Lobe.

Since mass transfer from the secondary is unstable, the white dwarf
breaks up on a dynamical timescale. After accreting some mass, the
primary is assumed to ignite helium and evolve to become a yellow 
supergiant with a helium-rich surface.
We assume conservation of angular momentum to compute the initial 
angular momentum distribution in a collisionless disk and subsequently 
in the giant envelope. At the end of shell-helium burning, the giant
contracts to form a white dwarf. We derive the surface rotation
velocity during this contraction. 

The calculation is repeated for a range of initial mass ratios, and
 also for the case of mergers between two helium white
 dwarfs; the latter will contract to the helium main-sequence rather than the
 white dwarf sequence. 

Assuming complete conservation of angular momentum, we predict
acceptable angular rotation rates for cool giants and during the initial subsequent
contraction. However such stars will only survive spin-up to reach the white dwarf
sequence (CO+He merger) if the initial mass ratio is close to unity. 
He+He merger products {\it must} lose angular momentum in order to
reach the helium main sequence. 

Minimum observed rotation velocities in extreme helium stars are lower
than our predictions by at least one half, indicating that CO+He
mergers must lose at least one half of their angular momentum,
possibly through a wind during shell-helium burning, but more likely
from the disk following secondary disruption. 
\end{abstract}


\section{Introduction}              
\label{intro}

Following \citet{Web84}, \citet{Sai02} have demonstrated that the most probable
  origin for extreme helium stars is a stellar merger in a binary system
  containing a carbon-oxygen and a helium white dwarf, although some
  may originate in systems containing two helium white dwarfs
  \citep{Sai00}. In the first case, the product ignites helium in a
  shell at the surface of the CO core and expands to become a
  helium-rich supergiant. After the helium-burning shell burns
  outwards through
  most of the helium-rich surface layers, the star contracts to become  
  a white dwarf. In the second case, the helium-ignition again occurs
  in a shell at the core-envelope boundary, but then burns inwards,
  lifting the electron-degeneracy in the helium core. When the
  helium-burning flame
  reaches the centre, the star essentially becomes a low-mass helium
  main-sequence star. 

  A criticism occasionally heard (but, to our knowledge, not yet written)
  is that the products of such mergers should have a very
  high angular momentum and thus should be observed as rapid rotators.
  While general arguments 
  suggest that this should not be a problem, it seemed appropriate to
  investigate the question more rigorously.

Previously \citet{Sai00, Sai02} examined the evolution of the internal 
structure of stars following two types 
of white dwarf merger. In this paper we focus on the 
evolution of the angular
momentum distribution from binary progenitor to final white
dwarf. We aim to set limits on the predicted rotation velocities of
the merger
products. 

\section{Angular momentum: general considerations}

The final mass of systems that become extreme helium
stars does not appear to exceed 0.9 $M_{\sun}$
\citep{Sai02}. Assuming tidal locking for the progenitor binary
we conclude that the orbital angular momentum is dominant. We already know
that angular momentum will be removed by gravitational
radiation on a timescale that depends on the initial separation of the
binary and will lead to a decrease in the orbital separation
\citep{Cha78}. For binary white dwarfs that
have a period of a few hours and an initial separation of $\sim
10^{11} {\rm cm} \sim 2 \Rsolar$ the time needed to merge is less than
the Hubble time. So
we expect a significant fraction of existing close white dwarf
binaries to merge. Because of orbital decay, at some
point the secondary fills its Roche Lobe.  The orbital period
at this point can be determined from the separation and is 
approximately 3 minutes.
Subsequently, mass starts to transfer
from the secondary to the primary. The transferred mass cannot all be
accreted onto the primary because of the Eddington limit. 
The mass transfer is unstable because of the inverse mass-radius
relation for white dwarfs. So the secondary will break up and 
form a thick disk round the primary on a dynamical timescale. 
For such systems, this is very short and of the order of the orbital
period.

To begin with, we assume conservation of total mass. This has been
demonstrated for the early evolution of white dwarf mergers
\citep{Seg97,Gue04}, but not for the later stages of their
evolution. 
In order to determine the characteristics of the angular
momentum profile of the disk we use two quantitites,
specific angular momentum and angular momentum density, defined in
section 3. The conservation of total angular momentum is a plausible
assumption. In addition, we assume conservation of angular momentum
for every single particle (or fluid element). Thus the
discussion reduces to an idealised collisionless model. 

This is a conservative assumption, which effectively reduces to a
worst case scenario. In the collisionless model, the transfer of 
angular momentum is slow compared to the dynamical timescale, so, it does not
have important effects. The opposite assumes a model in which
collisions are important, producing a viscous disk. The problem of
angular momentum transport in a viscous disk has been approached
already \citep{Lyn74}, where it was shown that ``the angular momentum
is steadily concentrated onto a small fraction of the mass which
orbits at greater and greater radii while the rest is accreted onto
the central body'' on an approximately dynamical timescale.
The problem for the viscous disk approximation is that the 
accretion rate for the surviving white dwarf then becomes very high
and it is not clear whether subsequent evolution will lead to the
assimilation of this mass or its complete ejection by radiation pressure. 
Consequently, we have started with the collisionless model to see
whether it, too, faces major obstacles.

Using the above assumption we propose an
angular momentum distribution profile for the disk.
In the purely collisionless model,  as will be shown,  
the inner radius of the disk is approximately twice the radius of 
the primary. The intervening gap will prevent accretion. 
In real systems, random collisions will broaden the angular momentum profile.
We will simulate this by applying a Gaussian smoothing function, where
the width of the Gaussian may be considered representative 
of the frequency of the collisions. Note that this procedure
does not change the total angular momentum of the system. 

Since the disk was created from the
decomposition of the secondary, which was a helium white dwarf, helium
will be its dominant constituent. Following \citet{Sai02}, after a small quantity
($0.004\Msolar$) of helium is
accreted by the primary, helium ignition occurs at the base of the
accreted layer. This energy source forces the star to expand to become
a giant. Models  indicate that the radius
of the giant will be two orders of magnitude greater than that of the disk, 
initially about 60 $R_{\sun}$. 

A simple model for the angular momentum distribution in the giant 
can be obtained by assuming that each cylindrical element in the disk 
forms a spherical shell conserving its angular momentum. In order to describe the
density profile of the giant we assume it consists of a degenerate
core, that is the carbon-oxygen white dwarf, and a convective
envelope. In the fully conservative case, the final mass of the 
convective envelope is equal to the mass of the initial disk, and hence 
of the helium secondary. 
Therefore we can describe an angular velocity distribution for the
star. 

Following calculation of the angular velocity distribution in
the giant merger product, we investigate how contraction affects the 
rotation. After helium-burning is completed, the stars will contract, 
and hence rotate more quickly. The question is whether their rotation
will approach the critical breakup velocity and what the consequences
might be. 

For the CO+He WD merger, we investigate three possible cases. 
In the first
case, the central region rotates as a rigid body and the angular
velocity profile of the envelope depends on the initial conditions,
meaning the disk angular momentum distribution. In the second case,
25\% of the convection zone near the surface rotates as a rigid body
as well as the central region whereas there is differential rotation
in the intermediate region. Thirdly, we examine the case of
completely rigid body rotation. This should be the ultimate equilibrium state
since no shear torques occur that could lead to angular
momentum transfer \citep{Lyn74, Pri81},
however it is not likely to be achieved, since it is a slow
process. 

Using the above assumptions we will model the angular momentum
evolution of a binary which consists initially of two white dwarfs.
 For the structure of the white dwarfs, 
we will adopt the models   described by \citet{Cha58}.
For the principal calculation we will consider the primary to be 
a carbon oxygen (CO) white dwarf with mass 0.6 $M_{\sun}$ and radius 
0.013$R_{\sun}$ and the secondary a helium (He) white dwarf with mass 
0.3 $M_{\sun}$ and radius 0.021 $R_{\sun}$ \citep{Ven95, Pan00}. In
addition we will consider CO+He binaries with masses of 
$0.7M_{\sun}+0.2M_{\sun}$ and $0.5M_{\sun}+0.4M_{\sun}$, 
in order to see the dependence, if any, of the angular velocity 
on the initial mass ratio. We also compute
appropriate quantities for a number of He+He white dwarf
configurations. Finally, the results of these calculations are compared
with observed angular velocities in extreme helium stars.

\section{Orbital Decay to Roche Lobe}
\label{Orbital Decay to Roche Lobe}

According to the General Theory of Relativity, two orbiting masses $M_{1}$,
$M_{2}$ with a separation $\alpha$ will radiate angular
momentum at a rate \citep{Lan58}

\begin{eqnarray}
\frac{\dot{J}}{J}=-\frac{32}{5}\frac{G^{3}}{c^{3}}\frac{M_{1}M_{2}M}{\alpha^{4}}
\end{eqnarray}

where $G$ is the gravitational constant, $c$ the speed of light,
and $M=M_{1}+M_{2}$ is the total mass of the system.  
Due to angular momentum loss their orbits will decay. 
The total angular momentum of a tidally locked system will be

\begin{eqnarray}
J_{\rm tot}=J_{1}^{o}+J_{1}^{s}+J_{2}^{o}+J_{2}^{s}
\end{eqnarray}

where the superscripts o and s refer to orbital and spin angular
momentum respectively and the subscripts 1 and 2 refer to the primary
and to the secondary. We can express the orbital angular momentum in the frame
of reference of the centre of mass

\begin{eqnarray}
J^{o}_{\rm tot}=M_{1}M_{2}\sqrt{\frac{G\alpha}{M}}
\end{eqnarray}

where M is the total mass of the system. Spin angular momentum can be expressed

\begin{eqnarray}
J^{s}_{\rm tot}=(I_{1}+I_{2})\omega 
\end{eqnarray}

where $\omega$ is the angular velocity and $I$ refers to the momentum
of inertia. Moments of inertia can be evaluated from the density
profile of the stars. However some distortions may occur that alter it
slightly \citep{Jam64, Tas78}. The
angular velocity is given by

\begin{eqnarray}
\omega=\sqrt{\frac{GM}{\alpha^{3}}}
\end{eqnarray}

Substituting angular momentum expressions from equations 3, 4
and differentiating with respect to $\alpha$, we obtain the following
expression

\begin{eqnarray}
\frac{\dot{J}}{J} = \left( \frac{M_{1}M_{2}}{2(M\alpha)^{1/2}} 
      - \frac{3I_{\rm tot}M^{1/2}}{2\alpha^{5/2}} \right ) \times\nonumber &&\\
   \left( \frac{M_{1}M_{2}\alpha^{1/2}}{M^{1/2}}
      - \frac{I_{\rm tot}{M^{1/2}}}{\alpha^{3/2}} \right )^{-1}\dot{\alpha}
\end{eqnarray}

As the orbit decays, the equipotential surface surrounding the two
stars (Roche lobes) shrinks until one component exactly fills its own
lobe. The Roche lobe radius is given by  \citep{Egg83}

\begin{eqnarray}
\alpha_{L}=\frac{0.49q^{2/3}}{0.6q^{2/3}+\ln(1+q^{1/3})}, 0<q<\infty
\end{eqnarray}
 
where $q$ is the ratio of the mass of the primary over the mass of the
secondary. 

For the $0.6+0.3{\rm M}_{\sun}$ system defined 
above, the secondary will fill its Roche lobe when the orbital 
separation is 0.067 ${\rm R}_{\sun}$. From eqns. 6 and 1 we can
estimate the orbital decay timescale. Assuming
an initial separation of 1.5 ${\rm R}_{\sun}$, which corresponds to an
orbital period of 5 hours, we obtain a timescale of $4.10^{9}$y, which
is less than a Hubble time. 
We note that observations of an increasing number of such close
binary white dwarfs \citep{Pac90, Mar95, Nap05}  are
commensurate with estimated merger rates \citep{Ibe96, Nel01}

\section{Disk Formation}

\label{Disk Formation}

When the secondary fills its Roche Lobe it disintegrates and its remnants
form a disk. We will use the assumption of angular momentum conservation 
to determine the mass distribution in the disk. First we
construct two useful quantities for the distribution of
angular momentum in the secondary. The first is the angular
momentum per unit mass, hereafter specific angular momentum, which will
be expressed as $dJ(m)/dm$, assuming the system shows cylindrical symmetry.
Since we generally assume a collisionless
model, this quantity remains constant throughout the problem unless
some mass is lost. The other quantity is the angular momentum
contained within a distance $r$ from the centre of the system,
hereafter angular momentum density, and will be expressed as
$dJ(r)/dr$. It depends strongly on the geometry of the
problem. When the secondary fills its Roche Lobes, it expresses the
angular momentum contained in a thin slice of the star at distance
$r$ from the rotation axis of the primary. 
The slice lies perpendicular to the radial
vector that points from the centre of the primary to the centre of the 
secondary. In the case of the disk, $dJ/dr$ expresses the angular momentum
carried by a thin cylindrical shell of radius $r$. In the
case of the giant star, it expresses the angular momentum carried by a
spherical shell at distance $r$ from the rotation axis of the star.

From the definition of angular momentum a particle of mass $dm$ that
lies at distance $r$ from the primary and moves at angular velocity $\omega$ has angular momentum

\begin{eqnarray}
dJ=r^{2}\omega dm 
\end{eqnarray}

\begin{figure}
\begin{center} 
\epsfxsize=8.7cm 
\epsfbox{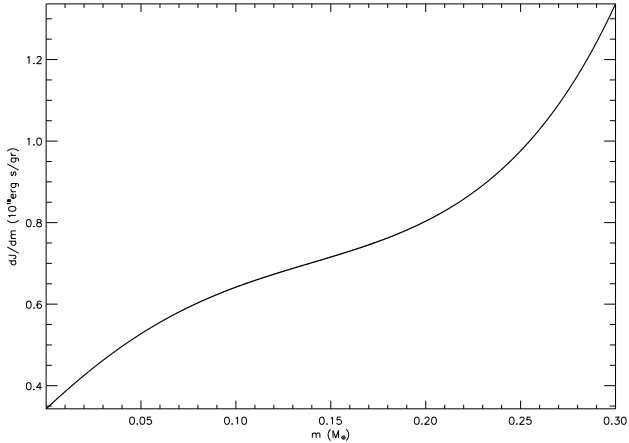}
\caption[Specific angular momentum distribution]
{The specific angular momentum $dJ/dm$. It remains constant throughout
  the problem, since we examine a collisionless model.}
\label{spec_am}
\end{center}
\end{figure}

\begin{figure}
\begin{center} 
\epsfxsize=8.7cm 
\epsfbox{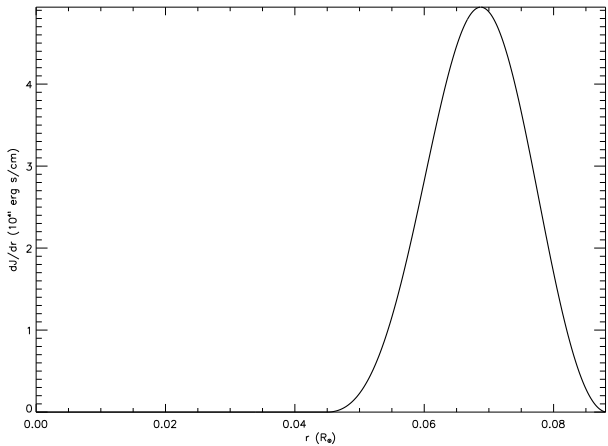}
\caption[Angular momentum density of secondary]{The angular momentum density $dJ/dr$ at the Roche
Lobe stage. We can see that most angular momentum is carried at a
distance slightly further from the centre of the secondary
(0.067$R_{\sun}$). 
This is because $dJ/dr$ is proportional to the mass
contained inside a slice of the star and to the square of the
distance. This product has a maximum at 0.069$R_{\sun}$ from the centre
of the primary.  }
\label{star_am_dens}
\end{center}
\end{figure}

Integrating the specific angular momentum over mass and the angular
momentum density over distance we obtain the total angular momentum of
the system. For our example conditions 
it is $4.5 \times 10^{50}{\rm \,erg\,s}$ and remains constant
throughout the evolution of the system, unless some mass is lost.
We can easily obtain from the orbital elements the angular velocity of
the secondary when $\alpha=\alpha_L$ to be $\omega =0.0346{\rm
  \,rad\,s^{-1}}$. 
The density of the secondary can be approximated by a polynomial function that fits the
data given by \citet{Cha58} for the structure of white dwarfs. Thence
we can derive the
specific angular momentum (Fig.~\ref{spec_am}) and the angular momentum density
(Fig.~\ref{star_am_dens}) for the system immediately before disruption of the secondary.
 
After disruption of the secondary, a disk is formed. Since the
disk is supported by gravity we assume it to be Keplerian. Therefore
the angular velocity distribution can easily be determined to first order,
ignoring the self gravitating effects of the disk itself.

\begin{eqnarray}
\omega (r)=\sqrt{\frac{GM_{1}}{r^{3}}}
\end{eqnarray}

The angular momentum for a particle of mass $m$ performing circular
motion of radius $r$, is $J=m\omega r^{2}$. Particles that carry less
of the angular momentum at the binary stage lie at the front
surface of the star ({\it i.e.} face-on to the the primary). The
angular momentum of these particles will determine the inner radius of
the disk. In our case, this is $r_{\rm in}=0.022R_{\sun}$. We can use the same
argument for the outer radius of the disk. In that case we evaluate
the angular momentum of a particle at the rear surface of the
secondary. We conclude that it is $r_{\rm out}=0.305R_{\sun}$. If we take
into account the self gravitating effect of the mass stored in the
disk, the outer radius is decreased 30\% to $r_{\rm out}=0.203R_{\sun}$,
whereas the inner radius will be unaffected. This may be
  shown by a first-order approximation. Assume that a particle of unit mass
  carrying angular momentum $J$ lies in the gravitational field of 
  a mass M. At radius $r$ the
  attractive force is $GM/r^2$. Including the gravitational field of
  a disk of mass $m$, the central force on the test particle, now at
  radius $r'$, is $G(M+m)/r'^2$. Since $J$ is constant, 
  the ratio of radii (for $m=0.3M_{\sun}$, $M=0.6M_{\sun}$) is
  $r'/r=2/3$. In higher order, the geometry of the disk must be
  considered, but this does not affect the angular momentum calculation.

\begin{figure}
\begin{center} 
\epsfxsize=8.7cm 
\epsfbox{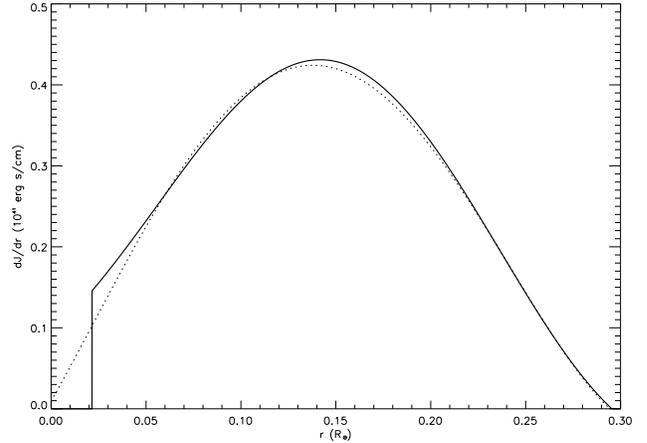}
\caption[Angular momentum density in disk]{The angular momentum density $dJ/dr$ of the
disk. Angular momentum is spread in a more extended region than it was
at the Roche Lobe stage. If we take into account the self gravitation of
the disk then its size decreases, but the total
angular momentum cannot change. Since we have a collisionless model, 
no mass is scattered closer to the star than $r_{\rm in} =
0.22{\rm R_{\sun}}$. 
The spin angular momentum of the primary has been neglected since it 
is three orders of magnitude smaller than the angular momentum of the
disk. The dotted line corresponds to the convolution of $dJ/dr$
with a Gaussian with $\sigma=0.005{\rm R_{\sun}}/\sqrt{2}$.
}
\label{disk_am_dens}
\end{center}
\end{figure}

Whether we use
the first or the second approximation for the angular momentum 
density will have no effect on our final results for the
rotation of the giant star, since the specific angular momentum does
not change. The choice only affects the angular momentum 
density at the disk stage (Fig.~\ref{disk_am_dens}).

During the formation of the disk collisions may take place. These will
broaden the angular momentum distribution, in particular to scatter
material into the region $r < r_{in}$.  
Angular momentum will also be transfered to outer parts of the disk
which will become more extended. In order to simulate the effect of
these collisions we convolve the angular momentum functions 
  ($dJ/dr$) 
with a
Gaussian of width $\sigma=0.005{\rm R_{\sun}}/\sqrt2$, 
where the width of the Gaussian may be taken to represent 
the overall efficiency of the collisions. The 
resulting density distribution demonstrates 
that only a very modest collisional redistribution of 
angular momentum is required to bring disk and star into contact 
(Fig.~\ref{disk_am_dens}). 

\section{Giant Stage}

\label{Giant Stage}

Following disk formation and providing gas has been scattered to $r <
r_{in}$, helium will be accreted from the disk onto the surface of the 
former primary, which becomes the degenerate core of the merged star. 
\citet{Sai02} adopted an accretion rate of roughly half the Eddington
rate.  After $\sim 0.029 M_{\sun}$ is accreted, a helium
flash occurs.  This leads the star to expand and form
a yellow giant. The expansion is rather rapid and lasts only 200
years. The radius of the giant  initially reaches $\sim 60 R_{\sun}$.
This stage of evolution may correspond to some
hydrogen-deficient carbon giants and the coolest of the R\,CrB stars
\citep{Sai02}. The radii of some models 
do exceed this value, so we surmise that the surface rotation 
will be lower than that deduced here. However, assuming a conservative
distribution of specific angular momentum, spindown during expansion
and spinup during contraction should lead to a similar final result.

Our objective is to predict the angular velocity profile for such a giant. In order to
model its density profile, we expect it to consist of a degenerate core
and a convective envelope. Using tables from \citet{Cha58},
  fitted with a $5^{\rm th}$ order poynomial, 
we approximate the envelope as a polytrope
of index $n=3/2$ and express the density as a function of $r$ that
fits the numerical results of the Lane-Emden equation for $n=3/2$
\citep{Moh80}. We assume that initially no angular momentum is
transferred between the 
various parts of the star.
Therefore we first make the approximation that the outside surface of the
star will form the 
outside shell of the star, whereas the inner parts of the disk will stay near the
core, and that there is no change in the specific angular
momentum. 

However, it is also possible that helium-shell ignition occurs before the 
disk has been fully accreted, so that the material which was accreted
first becomes the surface of the expanding giant, while the disk
survives for sometime {\it inside} the giant envelope. 
Subsequent accretion, or the disintegration of the disk, feeds mass 
into the interior of the giant, close to the core-envelope boundary. 
We will examine this case as well, 
assuming that the outer layer of the giant, with a mass equal to
the burning shell, 
comes from the inner part of the disk and the remainder 
of the envelope comes from the disentegration 
of the disk and obeys cylindrical symmetry.
  
We neglect differential rotation within a shell,
so a thin spherical shell of mass $dm$ at distance $r$ from the centre of the
star rotates as a rigid body with momentum of inertia

\begin{eqnarray}
I=\frac{2}{3}r^{2}dm 
\end{eqnarray}

The choice of cylindrical shells, rather than spherical, is made 
primarily on grounds of mathematical simplicity. There is some
evidence ({\it e.g.} the Sun) that rotation may be a function of
latitude -- possibly suggesting cylindrical symmetry. It will be
seen, however, that the difference between the extreme cases of 
rigid-body and differential rotation are sufficiently small that the 
choice of geometry in the latter are unlikely to be important.

\begin{figure}
\begin{center} 
\epsfxsize=8.7cm 
\epsfbox{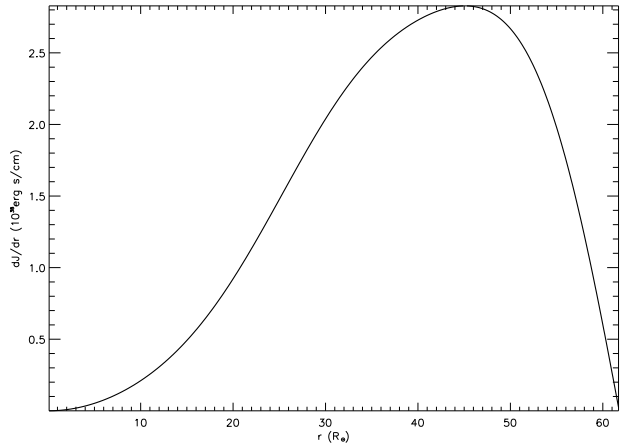}
\caption[]{The angular momentum density $dJ/dr$ of the convective envelope of the giant.}
\label{giant_am_dens}
\end{center}
\end{figure}

\begin{figure}
\begin{center} 
\epsfxsize=8.7cm 
\epsfbox{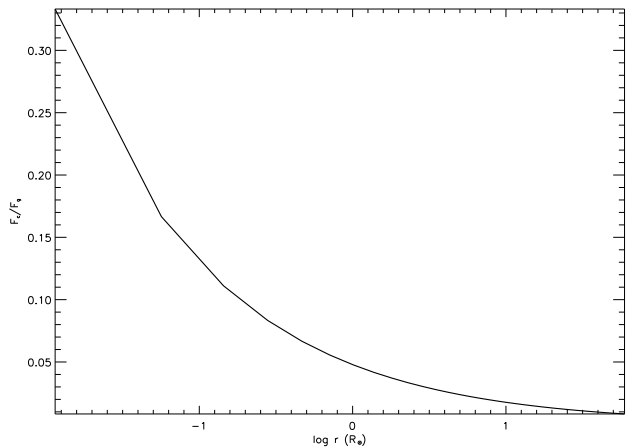}
\caption[]{The ratio of centrifugal force to gravitation force as a
function of radius for case 1. The plot shows only the differentially
rotating stellar envelope.}
\label{centrifugal}
\end{center}
\end{figure}

We examine a number of cases. 

\subsection*{Case 1: Rigid core + differential envelope}
\begin{figure}
\begin{center} 
\epsfxsize=8.7cm 
\epsfbox{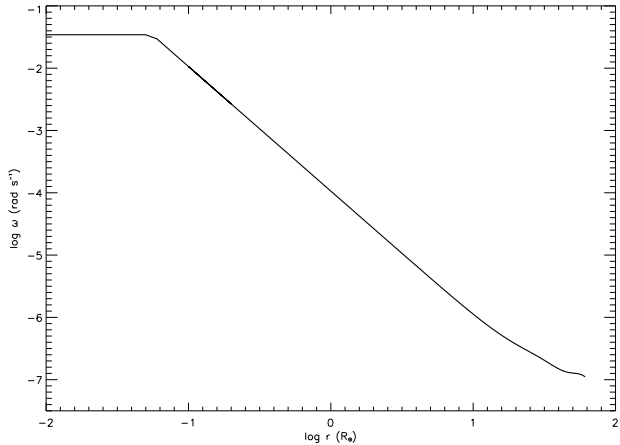}
\caption[]{Logarithmic plot of the angular velocity $\omega$ at the
  giant stage as a function of the distance from the rotation axis. 
  The central region rotates as a rigid body, whereas angular velocity 
  falls with distance from the axis.}
\label{case1_av}
\end{center}
\end{figure}

 The first case
involves two regions including a central region containing the degenerate core
and a shell of the star's envelope of width $0.045\,\Rsolar$ performing
rigid body rotation. Assuming that the core rotates with the
  original angular velocity of the primary white dwarf, this is the 
  critical radius
  for rigid-body rotation. Beyond this, the envelope is assumed
to rotate differentially
according to the angular momentum distribution determined by the disk
stage (Fig.~\ref{giant_am_dens}). 
The outer boundary to the region of solid body rotation is 
chosen in order to avoid the possibility of having a rotational
velocity 
that exceeds the orbital velocity at this point. 
This scenario predicts an angular velocity on
the outer layer of the star of $1.15\times 10^{-7}{\rm\,rad\,
s^{-1}}$, corresponding to an equatorial surface velocity 
$v_{\rm eq} = 4.95 {\rm\,
km\,s^{-1}}$. There is a slight discontinuity between the two regions
described above. 
It can be 
considered as a surface of infinite gradient $d \omega /dr$. 
In such a case the transfer 
of angular momentum will be very fast and the discontinuity will vanish. 
This will be the initial stage in
the giant's angular momentum evolution (Fig~\ref{case1_av}). 
Fig.~\ref{centrifugal} shows the ratio of centrifugal force to 
gravitation force as a function of radius. This drops rapidly outside
the core, so rotatation should not significantly affect the structure
of the envelope, although it may produce a slightly ellipsoidal star.

\begin{figure}
\begin{center} 
\epsfxsize=8.7cm 
\epsfbox{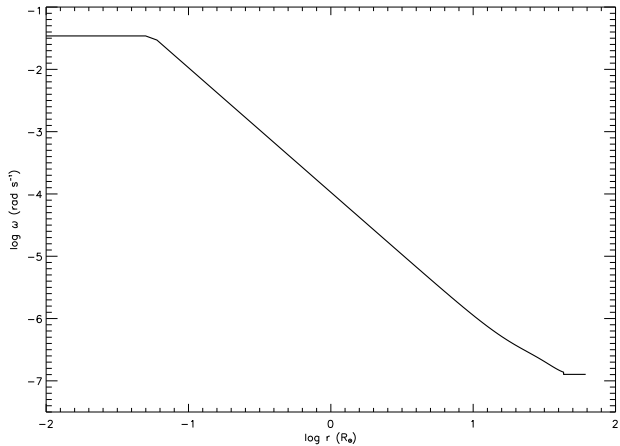}
\caption[]{The angular velocity $\omega$ of the disk, in the second
scenario. The central and the intermediate region rotate as in the
previous case (Fig.~\ref{case1_av}), however, there is a layer at the outter part
of the convective envelope that rotates as a rigid body. There are two
slight dicontinuities at the boundaries of the three regions.}
\label{case2_av}
\end{center}
\end{figure}

\subsection*{Case 2: Rigid core + transition layer + rigid envelope}
In the second case we divide the star into three regions, the first
is the core and a thin layer that rotates as a rigid body as before. 
The second region lies between $0.045\,\Rsolar$ and $45\,\Rsolar$, 
where the star rotates differentially as before. 
The third region lies between $45\,\Rsolar$ and the
surface. This boundary was chosen arbitrarily in order to represent a
case some way between the extremes discussed in cases 1 and 3.
We expect that near the surface, convection may be strong enough  that 
mixing will cause these outer layers to rotate as a rigid body 
(Fig.~\ref{case2_av}). 
Therefore the angular velocity for the surface and the outer
convection  
zone is $1.27\times 10^{-7}{\rm \,rad\,s^{-1}}$, corresponding to 
$v_{\rm eq} = 5.33 {\rm \,km\,s^{-1}}$. There is also a slight discontinuity 
at the $45\,\Rsolar$ boundary of $\delta \omega
=1.2\times 10^{-8}{\rm \,rad\,s^{-1}}$. This represents an intermediate
case, since angular momentum is transferred within the star. 

\subsection*{Case 3: Rigid body rotation}
The third case to consider is that of rigid body rotation for the
whole star. Shear torques may transfer angular momentum outwards 
from the core to the outer layers and thus make them move faster, while
the inner layers will rotate more slowly. 
Rigid body rotation is the equilibrium state since it is the
lowest energy state for an object with determined angular momentum
\citep{Lyn74, Pri81} and contains no shear torques that will 
transfer angular momentum. However, this equilibrium is not
  achieved by real stars even on very long timescales, 
 as illustrated by the Sun. It is unlikely to occur on the very
 short timescales and very low-density envelopes under consideration
 here. Nevertheless, it represents an important limiting case 
in which  most 
of the angular momentum is carried in the outer layers of the star. 
We can easily evaluate the moment of inertia for the
giant star by integrating Eqn.\,10 throughout the star to obtain
$I_{\rm giant}=2.85 \times 10^{2}{\rm
g\,cm^{2}}\,\Msolar^{-1}\,\Rsolar^{-2}$. From conservation of angular
momentum we find that $J_{\rm giant}=4.71 \times 10^{-5}{\rm \,erg\,
s}\,\Msolar^{-1}\,\Rsolar^{-2}$. Therefore

\begin{eqnarray}
\omega_{\rm giant}=\frac{J_{\rm giant}}{I_{\rm giant}}
\end{eqnarray}

\noindent which is $1.64\times 10^{-7}{\rm \,rad\,s^{-1}}$ 
corresponding to $v_{\rm eq} = 7.1 {\rm \,km\, s^{-1}}$. 
This case represents the fastest rotation of the
surface layers and sets an upper limit for the observed rotation
velocity.

We conclude that the equatorial surface velocities do not depend strongly on the
details of the model. Using the above argument we can define an upper
limit on the angular momentum that corresponds to rigid body
rotation. Any velocity observed should be less than this limit. We
have not taken into account any losses of angular momentum due to mass
ejection. Such phenomena will lead to lower angular momentum and
therefore lower angular velocity.  In the polytropic density profile we
have neglected any term from rotation; this gives a good approximation
since the linear velocity of circular motion due to the gravitational field
of the star at a radius equal to the giant star radius is
$53{\rm \,km\,s^{-1}}$, or about one order of magnitude
greater than the velocities we found. In addition, any
distortion caused by rotation will increase the moment of
inertia of the star and decrease the angular velocity, 
supporting the statement that rigid body rotation of a
spherical body represents the maximum possible equatorial surface
velocity.

\subsection*{Case 4: Rigid core $0.87{\rm M}_{\odot}$ + differential envelope}
The fourth case can be considered as a combination of cases 1 and 3.
During evolution as a giant, the helium-burning shell eats up
the envelope, so the core mass
increases, while the core radius remains virtually
unchanged. Conversely, the envelope retains the same radius but 
a drastically reduced mass. We consider a final configuration in which
the envelope mass is $0.03{\rm M}_{\odot}$. The moment
of inertia of the core increases but cannot exceed 
$1.4\times 10^{-5} {\rm \,g\,cm^2}$. Even in this case, 
no more than 15\% of the total angular momentum can be stored in the core
before the core reaches its breakup limit. Therefore the envelope 
will have to rotate about $30$ times faster than in case 1. 

Such rotation is extremely fast and seems implausible; at least, it
is not observed. The problem is that, to be conservative, angular momentum
must be transferred into the core at the same time as the envelope 
is ingested. However, once the core reaches breakup velocity the
transfer of angular momentum becomes impossible, so the momentum must 
stay in the envelope, creating a paradox. Interestingly, this
parallels the case for the viscous disk \citep{Lyn74} in which angular 
momentum is expelled outwards as mass migrates inwards. 

This calculation therefore strongly suggests that if stable
shell-burning giants are to be produced by white dwarf mergers, then a
large fraction of the total angular momentum must be dissipated from
the disk {\it before} it is accreted onto the primary. 

While this case is the most interesting and realistic of the four
    considered, it has presented a paradox for the transfer of 
    angular momentum from the envelope to the growing core. The 
    consequences of this deserve more attention than we are currently 
    equipped to explore.
 
\begin{figure}
\begin{center} 
\epsfxsize=8.7cm 
\epsfbox{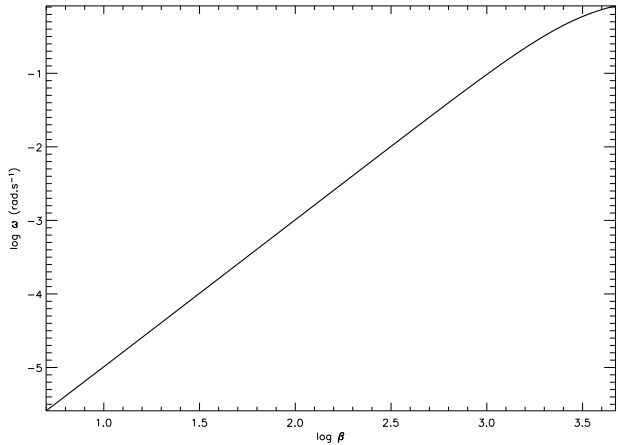}
\caption[]{Logarithmic plot of the angular velocity $\omega$ as the
  star contracts by a fraction $\beta=R_0/R$. $R_0$ is the radius of
  the giant before contraction. }
\label{ehe_omega}
\end{center}
\end{figure}

\begin{figure}
\begin{center} 
\epsfxsize=8.7cm 
\epsfbox{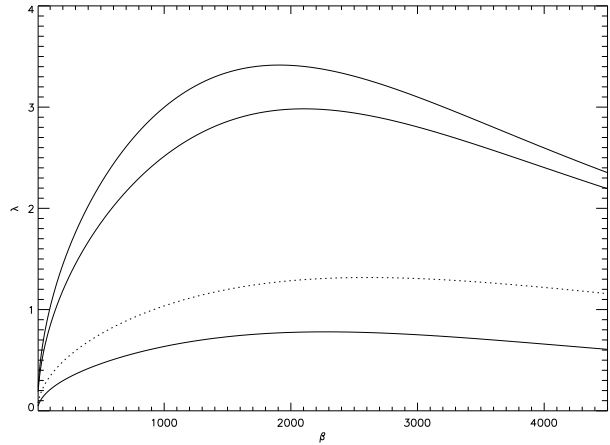}
\caption[]{The ratio of the velocity due to rotation over the velocity
of the circular orbit on the surface of the star $\lambda$, as a
function of $\beta$ which is the contraction factor for three pairs of
masses of the progenitors, from top to bottom
$0.7M_{\sun}+0.2M_{\sun}$, $0.6M_{\sun}+0.3M_{\sun}$ and
$0.5M_{\sun}+0.4M_{\sun}$. Contraction cannot continue if $\lambda$
approaches unity. In order to become a white dwarf the star needs to
contract by a factor of about $5.10^{3}$. We can see that for progenitor masses
of $0.5M_{\sun}+0.4M_{\sun}$, contraction can reach the white dwarf radius,
whereas for the other two cases angular momentum needs to be removed.
The dotted line corresponds to the model of a merger with 
progenitor masses of  $0.6M_{\sun}+0.3M_{\sun}$ where 50\% of the secondary 
mass is lost.
}
\label{rot_orb_v}
\end{center}
\end{figure}

\section{Homologous Contraction}

\label{Homologous Contraction}

Having considered the angular momentum distribution of a star as it
evolves from a white dwarf binary through to being a giant, we next
consider the angular velocity distribution following shell helium 
extinction and the subsequent contraction at constant luminosity 
towards the white dwarf cooling track.  

Since the core is degenerate, we need only consider the evolution of
the envelope, which we shall consider to contract
homologously. A simple model can be developed by assuming that the 
star rotates as a rigid body (case 3). 

However, no crucial differences will be found if we assume differential
rotation, as demonstrated in the previous section. The only effect
would be to find lower surface equatorial velocities. In order to become a
white dwarf the star needs to reduce in size by a factor  $\sim2
\times 10^{-4}$. 
Leaving the core unaffected, while the star shrinks, we
evaluate the angular velocity and the equatorial surface velocity 
corresponding to this rotation. We plot the angular velocity  
(Fig.~\ref{ehe_omega}) and the ratio of the
linear equatorial rotation velocity $v_{\rm eq}$ to the linear velocity of a particle
moving on a circular orbit $v_{\rm orb}$ at the surface: 

\begin{eqnarray}
\lambda = v_{\rm eq} / v_{\rm orb}
\end{eqnarray}

\noindent (Fig.~\ref{rot_orb_v}: middle curve). The quantity on the horizontal axis is the ratio
$\beta$ of the radius of the giant $R_0$ to the radius during
contraction $R$. The momentum of inertia decreases with the square of the
radius, which will lead the star to rotate faster
(Fig.~\ref{ehe_omega}). Note that contraction implies increasing $\beta$.

Problems occur when the rotation velocity approaches or exceeds the value of
velocity for a circular orbit ($\lambda\gtrsim1$).
Thus, when the star contracts to $\sim0.007$ times its initial
radius, the two velocities become equal and the star cannot shrink
anymore unless some angular momentum is lost. At this point the
velocity is of order $\sim 10^{2}{\rm \,km\,s^{-1}}$. This is much lower than the
speed of light and relativistic effects may be ignored.

\begin{table*}
 \caption{Theoretical predictions for the rotation of helium main-sequence
   stars produced by He+He white dwarf mergers for various combinations of initial
   masses.}
 \label{tab_he+he} 
\begin{center}
 \begin{tabular}{ccccccc}
  \noalign{\smallskip}
 \noalign{\smallskip}
\hline
 \noalign{\smallskip}
 \noalign{\smallskip}
   $ M /{\rm M_{\sun}}$
 & $M_{1}+M_{2} /{\rm M_{\sun}}$
 & $I / 10^{-3} {\rm M_{\sun} R_{\sun}^{2}}$
 & $J / 10^{-5} {\rm erg\,s M_{\sun}^{-1} R_{\sun}^{-2}}$ 
 & $\omega / {\rm rad\,s^{-1}}$
 & $v_{\rm rot} / 10^{3} {\rm km\,s^{-1}}$ 
 & $v_{\rm orb} / 10^{3} {\rm km\,s^{-1}}$ \\
 \hline  
 \noalign{\smallskip}
 0.5 & 0.3+0.2 & 1.16 &  2.53 &  0.044  & 3.2 & 0.95 \\ 
0.5 & 0.4+0.1 & 1.16 &  1.71 &  0.030  & 2.2 & 0.95 \\
0.7 & 0.4+0.3 & 3.18 & 3.83 & 0.012 & 1.2 & 0.96  \\
0.7 & 0.5+0.2 & 3.18 & 3.18 &  0.010 & 1.0 & 0.96 \\
\noalign{\smallskip}
\hline
\noalign{\smallskip}
\end{tabular}
\end{center}
\end{table*}

\section{An angular momentum problem?}

So far we have examined a fully conservative scenario from the Roche
lobe stage and after for a $0.6+0.3{\rm M}_{\sun}$ binary.  However, 
much energy is
ejected at the breakup of the secondary. Some of this energy may be
deposited in the disk as thermal kinetic energy, while some material of high
angular momentum is likely to leave the system, although the maximum possible
mass lost has been shown to be small \citep{Han99}.

We therefore examine 
whether the contraction of the giant can proceed conservatively 
if some mass (and hence its associated specific angular momentum) 
is ejected.
Although this is contrary to our initial assumptions and previous 
numerical results \citep{Seg97,Han99,Gue04}, this mass ejection is
treated as taking place during the phase of disk formation. This 
approximation avoids numerical complications which are beyond 
the scope of this paper, but allows us to obtain a rough estimate 
of the resulting configurations when mass loss is taken into account.

 We have solved 
the problem for the cases where 
1\%, 5\%, 10\%, 25\% and 50\% of the mass of the secondary is ejected
during the merger process. In all cases, the giant cannot contract
to the radius of a white dwarf. 
Fig.~\ref{rot_orb_v} includes the case in which 50\% of
the mass of the secondary is ejected. In this test case, 2/3 of the 
total angular momentum is removed, since
particles carrying more angular momentum are more likely to escape.
$\lambda$ reaches unity when $\beta \approx 1000$, which means that
the star can reach a small size ($10^{-3}$ times its 
original), but not less than five times the
required white dwarf radius.

In order to investigate the r\^ole of the initial conditions, 
we have solved the problem for various relative masses of the
white dwarf binary components. We have plotted the ratio of the 
equatorial rotation velocity to the surface orbital velocity ($\lambda$)
for progenitor binaries of $0.7+0.2{\rm M}_{\sun}$ and
$0.5+0.4{\rm M}_{\sun}$ in addition to the test case of 
$0.6+0.3{\rm M}_{\sun}$ (Fig.~\ref{rot_orb_v}). When
  $\lambda\gtrsim1$, contraction cannot proceed. We see that
for mass ratios nearer unity the star {\it can} contract to white dwarf size
without the need for additional angular momentum loss. This is expected, since
such systems need to lose more angular momentum before the secondary star fills its
Roche Lobe. This is because the white dwarf radius varies inversely
  with mass, so more massive white dwarfs will have smaller radii and
  will fill their Roche lobes later. Close orbits involve less angular
momentum of course, and their products will rotate more slowly.

Therefore conservation of angular momentum does present a problem
for the white dwarf merger model in most, but not, all cases. The
problem arises either during helium-shell burning, when angular
momentum has to be transferred to the envelope to avoid the core
reaching breakup velocity, or during contraction, when the surface
reaches breakup velocity.

\section{He+He white dwarf mergers}

The case of merger between two helium white dwarfs is initially
identical to that of the CO+He merger \citep{Sai00}. After shell 
helium ignition,
the star expands to become a giant, and the envelope is capable of
storing substantial angular momentum. However, the star evolves to
become a helium main-sequence star on a short timescale
\citep{Ibe90}. 
The case is 
analogous to the CO+He merger which must spin up during contraction 
to become a white dwarf.

We have calculated rotation rates for helium main-sequence stars
arising from a number of initial 
configurations.
The final configuration is an $n=5/2$ polytrope 
with overall dimensions given by \citet{Pac71}.

Table \ref{tab_he+he} shows the run of the moments of inertia $I$, 
total angular momenta $J$, and angular and linear equatorial rotation 
velocities ($\omega, v_{\rm rot}$) for
a series of inital configurations ($M_1,M_2$). The predicted values for 
$v_{\rm rot}$ are very high and, in most cases, substantially 
exceed the critical orbital velocity at the surface ($v_{\rm orb}$). 
Clearly such stars cannot exist and it is necessary to find a
mechanism for 
expelling angular momentum at an earlier stage. 
In this case losing some angular momentum at the
breakup of the secondary when it fills its Roche Lobe is likely, since
the total mass of the binary and the binding energy is lower. 
In addition to this, less massive white dwarfs have
larger radii and fill their Roche lobe earlier. This will
enable mass carrying angular momentum to escape the system earlier.

\section{Observed Rotation Rates}

\begin{figure}
\begin{center} 
\input{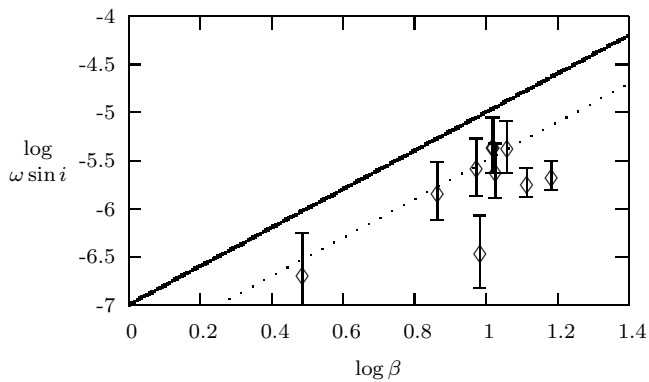}
\caption[]{Projected angular rotation $\log \omega \sin i$ as a function of
  relative inverse radius $\beta$ for extreme helium stars. The dotted
  line represents a homologous contraction, i.e. $\omega \propto
  1/R^2$, and the solid line represents the maximum theoretical
  rotation rate shown in Fig.~\ref{ehe_omega}. }
\label{fig_wr}
\end{center}
\end{figure}

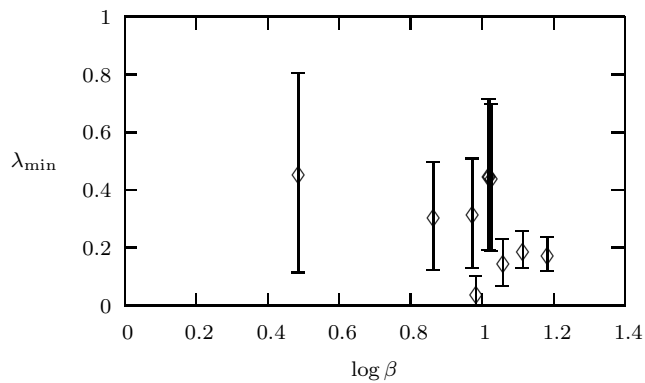
\begin{figure}
\begin{center} 
\setlength{\unitlength}{0.240900pt}
\ifx\plotpoint\undefined\newsavebox{\plotpoint}\fi
\begin{picture}(1027,616)(0,0)
\sbox{\plotpoint}{\rule[-0.200pt]{0.400pt}{0.400pt}}%
\put(181.0,123.0){\rule[-0.200pt]{4.818pt}{0.400pt}}
\put(161,123){\makebox(0,0)[r]{ 0}}
\put(947.0,123.0){\rule[-0.200pt]{4.818pt}{0.400pt}}
\put(181.0,214.0){\rule[-0.200pt]{4.818pt}{0.400pt}}
\put(161,214){\makebox(0,0)[r]{ 0.2}}
\put(947.0,214.0){\rule[-0.200pt]{4.818pt}{0.400pt}}
\put(181.0,305.0){\rule[-0.200pt]{4.818pt}{0.400pt}}
\put(161,305){\makebox(0,0)[r]{ 0.4}}
\put(947.0,305.0){\rule[-0.200pt]{4.818pt}{0.400pt}}
\put(181.0,395.0){\rule[-0.200pt]{4.818pt}{0.400pt}}
\put(161,395){\makebox(0,0)[r]{ 0.6}}
\put(947.0,395.0){\rule[-0.200pt]{4.818pt}{0.400pt}}
\put(181.0,486.0){\rule[-0.200pt]{4.818pt}{0.400pt}}
\put(161,486){\makebox(0,0)[r]{ 0.8}}
\put(947.0,486.0){\rule[-0.200pt]{4.818pt}{0.400pt}}
\put(181.0,577.0){\rule[-0.200pt]{4.818pt}{0.400pt}}
\put(161,577){\makebox(0,0)[r]{ 1}}
\put(947.0,577.0){\rule[-0.200pt]{4.818pt}{0.400pt}}
\put(181.0,123.0){\rule[-0.200pt]{0.400pt}{4.818pt}}
\put(181,82){\makebox(0,0){ 0}}
\put(181.0,557.0){\rule[-0.200pt]{0.400pt}{4.818pt}}
\put(293.0,123.0){\rule[-0.200pt]{0.400pt}{4.818pt}}
\put(293,82){\makebox(0,0){ 0.2}}
\put(293.0,557.0){\rule[-0.200pt]{0.400pt}{4.818pt}}
\put(406.0,123.0){\rule[-0.200pt]{0.400pt}{4.818pt}}
\put(406,82){\makebox(0,0){ 0.4}}
\put(406.0,557.0){\rule[-0.200pt]{0.400pt}{4.818pt}}
\put(518.0,123.0){\rule[-0.200pt]{0.400pt}{4.818pt}}
\put(518,82){\makebox(0,0){ 0.6}}
\put(518.0,557.0){\rule[-0.200pt]{0.400pt}{4.818pt}}
\put(630.0,123.0){\rule[-0.200pt]{0.400pt}{4.818pt}}
\put(630,82){\makebox(0,0){ 0.8}}
\put(630.0,557.0){\rule[-0.200pt]{0.400pt}{4.818pt}}
\put(742.0,123.0){\rule[-0.200pt]{0.400pt}{4.818pt}}
\put(742,82){\makebox(0,0){ 1}}
\put(742.0,557.0){\rule[-0.200pt]{0.400pt}{4.818pt}}
\put(855.0,123.0){\rule[-0.200pt]{0.400pt}{4.818pt}}
\put(855,82){\makebox(0,0){ 1.2}}
\put(855.0,557.0){\rule[-0.200pt]{0.400pt}{4.818pt}}
\put(967.0,123.0){\rule[-0.200pt]{0.400pt}{4.818pt}}
\put(967,82){\makebox(0,0){ 1.4}}
\put(967.0,557.0){\rule[-0.200pt]{0.400pt}{4.818pt}}
\put(181.0,123.0){\rule[-0.200pt]{189.347pt}{0.400pt}}
\put(967.0,123.0){\rule[-0.200pt]{0.400pt}{109.369pt}}
\put(181.0,577.0){\rule[-0.200pt]{189.347pt}{0.400pt}}
\put(181.0,123.0){\rule[-0.200pt]{0.400pt}{109.369pt}}
\put(40,350){\makebox(0,0){$\lambda_{\rm min}$}}
\put(574,21){\makebox(0,0){$\log \beta$}}
\put(454.0,175.0){\rule[-0.200pt]{0.400pt}{75.402pt}}
\put(444.0,175.0){\rule[-0.200pt]{4.818pt}{0.400pt}}
\put(444.0,488.0){\rule[-0.200pt]{4.818pt}{0.400pt}}
\put(666.0,179.0){\rule[-0.200pt]{0.400pt}{40.712pt}}
\put(656.0,179.0){\rule[-0.200pt]{4.818pt}{0.400pt}}
\put(656.0,348.0){\rule[-0.200pt]{4.818pt}{0.400pt}}
\put(727.0,182.0){\rule[-0.200pt]{0.400pt}{41.435pt}}
\put(717.0,182.0){\rule[-0.200pt]{4.818pt}{0.400pt}}
\put(717.0,354.0){\rule[-0.200pt]{4.818pt}{0.400pt}}
\put(733.0,123.0){\rule[-0.200pt]{0.400pt}{11.081pt}}
\put(723.0,123.0){\rule[-0.200pt]{4.818pt}{0.400pt}}
\put(723.0,169.0){\rule[-0.200pt]{4.818pt}{0.400pt}}
\put(752.0,210.0){\rule[-0.200pt]{0.400pt}{57.334pt}}
\put(742.0,210.0){\rule[-0.200pt]{4.818pt}{0.400pt}}
\put(742.0,448.0){\rule[-0.200pt]{4.818pt}{0.400pt}}
\put(754.0,210.0){\rule[-0.200pt]{0.400pt}{57.334pt}}
\put(744.0,210.0){\rule[-0.200pt]{4.818pt}{0.400pt}}
\put(744.0,448.0){\rule[-0.200pt]{4.818pt}{0.400pt}}
\put(757.0,209.0){\rule[-0.200pt]{0.400pt}{55.648pt}}
\put(747.0,209.0){\rule[-0.200pt]{4.818pt}{0.400pt}}
\put(747.0,440.0){\rule[-0.200pt]{4.818pt}{0.400pt}}
\put(775.0,154.0){\rule[-0.200pt]{0.400pt}{17.586pt}}
\put(765.0,154.0){\rule[-0.200pt]{4.818pt}{0.400pt}}
\put(765.0,227.0){\rule[-0.200pt]{4.818pt}{0.400pt}}
\put(806.0,182.0){\rule[-0.200pt]{0.400pt}{13.972pt}}
\put(796.0,182.0){\rule[-0.200pt]{4.818pt}{0.400pt}}
\put(796.0,240.0){\rule[-0.200pt]{4.818pt}{0.400pt}}
\put(845.0,177.0){\rule[-0.200pt]{0.400pt}{13.009pt}}
\put(835.0,177.0){\rule[-0.200pt]{4.818pt}{0.400pt}}
\put(454,332){\raisebox{-.8pt}{\makebox(0,0){$\Diamond$}}}
\put(666,264){\raisebox{-.8pt}{\makebox(0,0){$\Diamond$}}}
\put(727,268){\raisebox{-.8pt}{\makebox(0,0){$\Diamond$}}}
\put(733,143){\raisebox{-.8pt}{\makebox(0,0){$\Diamond$}}}
\put(752,329){\raisebox{-.8pt}{\makebox(0,0){$\Diamond$}}}
\put(754,329){\raisebox{-.8pt}{\makebox(0,0){$\Diamond$}}}
\put(757,325){\raisebox{-.8pt}{\makebox(0,0){$\Diamond$}}}
\put(775,191){\raisebox{-.8pt}{\makebox(0,0){$\Diamond$}}}
\put(806,211){\raisebox{-.8pt}{\makebox(0,0){$\Diamond$}}}
\put(845,204){\raisebox{-.8pt}{\makebox(0,0){$\Diamond$}}}
\put(835.0,231.0){\rule[-0.200pt]{4.818pt}{0.400pt}}
\put(181.0,123.0){\rule[-0.200pt]{189.347pt}{0.400pt}}
\put(967.0,123.0){\rule[-0.200pt]{0.400pt}{109.369pt}}
\put(181.0,577.0){\rule[-0.200pt]{189.347pt}{0.400pt}}
\put(181.0,123.0){\rule[-0.200pt]{0.400pt}{109.369pt}}
\end{picture}
\caption[]{Projected angular rotation as a fraction of breakup
  velocity $\lambda$ as a function of 
  relative inverse radius $\beta$ for extreme helium stars. }
\label{fig_lr}
\end{center}
\end{figure}

\begin{table*}
 \caption{Projected rotation velocities for extreme helium stars.}
 \label{tab_obs}
 \begin{center}
 \begin{tabular}{lccccccl}
  \noalign{\smallskip}
 \noalign{\smallskip}
\hline
\noalign{\smallskip}
 Star               & $T_{\rm eff}/{\rm K}$ & $\log g$ & $v_{\rm eq} \sin
 i/{\rm km\,s^{-1}}$&$R$/R$_{\sun}$& $\omega \sin i/10^6 {\rm rad\,s^{-1}}$ & Reference  \\
\hline  
\noalign{\smallskip}
FQ Aqr             &$8750\pm300$ & $0.3\pm0.3$  &$20\pm5$&$136\pm96$&$0.22\pm0.17$& \citet{Pan06} \\   
HD168476           &$13500\pm500$& $1.6\pm0.25$ &$25\pm5$&$ 23\pm13$&$1.6\pm1.0$& \citet{Pan06} \\
LSS\, 99           &$15330\pm500$& $1.9\pm0.25$ &$30\pm5$&$ 16\pm 9$&$2.9\pm1.7$& \citet{Jef98} \\ 
HD124448           &$15500\pm500$& $1.9\pm0.25$ &$4\pm5$ &$ 16\pm 9$&$0.38\pm0.5$& \citet{Pan06} \\
LSS\,4357          &$16130\pm500$& $2.0\pm0.25$ &$45\pm5$&$ 14\pm 8$&$4.9\pm2.8$& \citet{Jef98} \\
LS\,II$+33^{\circ}5$&$16180\pm500$&$2.0\pm0.25$ &$45\pm5$&$ 14\pm 8$&$4.9\pm2.8$& \citet{Jef98} \\
V1920 Cyg          &$16300\pm900$& $1.7\pm0.25$ &$40\pm5$&$ 23\pm13$&$2.6\pm1.5$& \citet{Pan06}  \\
BD$+10^{\circ}2179$&$16900\pm500$& $2.55\pm0.2$ &$18\pm5$&$  6\pm 3$&$4.7\pm2.5$& \citet{Pan06}  \\ 
LSE\,78            &$18000\pm700$& $2.0\pm0.1$  &$20\pm5$&$ 15\pm 3$&$2.0\pm0.7$& \citet{Jef93a} \\
DY\,Cen            &$19500\pm500$& $2.15\pm0.1$ &$20\pm5$&$ 13\pm 3$&$2.4\pm0.8$& \citet{Jef93b} \\
\noalign{\smallskip}
\hline
\noalign{\smallskip}
\end{tabular}
\end{center}
\end{table*}

\label{Observations}

Observations suggest modest rotation velocities for extreme helium stars.
Measurements of the projected rotation velocity $v_{\rm eq} \sin i$
for ten objects have been obtained by fitting line profiles to absorption
lines in the optical spectrum (Table~\ref{tab_obs}). For small values
of $v_{\rm eq} \sin i$, these may been overestimated because
of the difficulty of deconvolving instrumental and rotation
broadening profiles in the observed spectra. Table~\ref{tab_obs} also
gives the measured values of $T_{\rm eff}$ and $\log g$. By assuming
a core-mass shell-luminosity ($M_{\rm c}-L_{\rm s}$) relation \citep[cf.]{Sai88}, 
the latter may be combined to estimate both a mass and a radius for
the extreme helium stars. Hence we can obtain the projected (minimum) angular
rotation velocity $\omega \sin i$. We can normalise the relative radii
by assuming a minimum $T_{\rm eff,0}$ before contraction at constant
luminosity and hence derive $\beta = R_0/R = (T_{\rm eff}/T_{\rm
  eff,0})^2$. 
We have adopted $T_{\rm eff,0}=5\,000 {\rm K}$, the value at which
post-merger giants commence their contraction \citep{Sai02}. A different
value would impose a horizontal offset onto Fig.~\ref{fig_wr}, which
demonstrates an overall increase in $\omega \sin i$ with increasing
$\beta$ (and hence with increasing $T_{\rm eff}$). It is interesting
that the maximum slope represented by these data corresponds to the
spin-up of a homologously contracting star, $\omega \propto R^{-2}$,
as theory predicts. 

The projected rotation rates are all smaller than the maximum rate
predicted by theory assuming complete conservation of angular momentum 
through the white dwarf merger. Assuming a random distribution of
inclination angles, the average rotation rates should be 
$\pi/4$ times the maximum rates. The observed rates are less than one third of
the maximum rate. This implies that at least half of the angular 
momentum must have been lost, either during the merger process or 
whilst the star was a cool giant, {\it e.g.} in a wind. 

The minimum rotation rate as a fraction of breakup velocity
$\lambda_{\rm min} = v_{\rm eq} \sin i/ v_{\rm orb} $ can also be
checked  using the same estimates for radius
(Fig.~\ref{fig_lr}). 
All objects are currently slow rotators ($v_{\rm eq} \sin i$ for
FQ\,Aqr is probably overestimated for the reasons given above). 

In other words, the observed rotation rates for extreme helium stars
are consistent with homologous contraction of the envelope at constant 
luminosity. They are not consistent with complete conservation of angular
momentum during the white dwarf merger. With angular velocities less
than one third the predicted rates, the likelihood of breakup as these
stars contract towards the white dwarf phase is much reduced compared
with Fig.~\ref{rot_orb_v}. 

\section{Conclusion}

Assuming that angular momentum is strictly conserved within the system,
we have calculated the rotation velocities of stars produced by the
merger of white dwarf binaries over a range of interesting initial
mass ratios and possible outcomes. These include evolution through the
giant phase and subsequently towards either the white dwarf or the
helium main-sequence. We have demonstrated that while it may be
possible to produce the initial giants, the star (or its core) 
will spin up as material is processed into the core or as the overall star 
contracts towards either of the compact configurations. This spinup
would be high enough to cause breakup. 

More critically, the rotation velocities predicted under total
conservation of angular momentum for post-giant
objects are greater than those observed in the putative CO+He merger
products, the extreme helium stars, by a factor $\gtrsim 3$. 

Therefore, for the merger model to work, the star {\it must} lose
angular momentum at some point in its evolution. The most likely time
for this to occur is following formation of the disk after breakup of
the secondary. In order to assume total conservation of angular
momentum, it was necessary to assume that this disk is collisionless,
but that assumption would also prevent the accretion of material onto the
primary. 

In fact the completely opposite
assumption must be considered. \citet{Lyn74} argue that, 
in a viscous disk, angular momentum will be transported outwards 
while mass is transported inwards, on the viscous timescale 
$\tau_{\rm visc} \approx R^2/\nu$, 
where $R$ is the disk radius and $\nu$ the viscosity. The latter is
approximately $\frac{1}{3}c_{\rm s}H$, where $c_{\rm s}$ is the sound
speed and the scale height 
$ H\approx R c_{\rm s}/v_{\rm c}$, 
$v_{\rm c}$ 
being the Keplerian velocity for a circular orbit at
radius $R$ having a period $P_{\rm R}$. Rearranging, we obtain 
 $\tau_{\rm visc} \sim 3R/v_{\rm c} (R/H)^2 $. If we take the disk
radius to be approximately four times the white dwarf radius $R \sim 4
R_{\rm wd}$, and the
disk scale height to be twice the white dwarf radius, we obtain 
$v_{\rm c} \approx 6R/P_{\rm R}$ and  $\tau_{\rm visc} \sim 2 P_{\rm
  R} \sim 500{\rm s}$.   
Such a mechanism provides a very efficient sink for the angular
momentum,
but poses a problem for the mass, which will be dumped onto the
surface of the white dwarf at a rate far in excess of the Eddington
rate ($\dot M_{\rm Edd}$). While accretion rates significantly higher
than $\dot M_{\rm Edd}$ may be accommodated in a nonspherical
geometry, {\it e.g.} by accretion at the equator balanced by 
radiation in the polar axis, the disparity here seems to be
insurmountable. Mass must be stored somewhere in the system where it is
not supported by hydrostatic forces. 

A hybrid solution might derive from a mechanism suggested first by 
\citet{Lyn74} and explored again by \citet{Pop91}. This considers
how material from an accretion disk would spin up the accretor (note, this
  is a true accretion disk, and not a disrupted secondary). When the
  accretor reaches critical velocity, it interacts with the disk,
  actually spinning up the disk. If this were to happen, then we could
  get momentum back into the disk, propagate it outwards and eject
  it from the outer edge of the disk.

A more detailed study of the dynamics of the binary disruption, 
disk formation and subsequent accretion is therefore warranted. 
Some progress has been made with smoothed particle hydrodynamics
\citep{Ben90,Seg97,Gue04}; calculations have dealt principally with
the dynamics of the secondary disintegration and subsequent 
disk formation, and give useful insight into the disk heating and mass
distribution, {\it i.e.} no explosion and negligible mass loss from the
system. However, none goes far enough to establish the long-term
outcome during and following disk accretion.

\section*{Acknowledgements}
K. Gourgouliatos is grateful
to the Armagh Observatory for a summer studentship during which this work was
carried out. Research at the Armagh Observatory is funded by the
Northern Ireland Department of Culture, Arts and Leisure. The authors
are grateful to Profs D. Lynden-Bell and M.~E. Bailey for stimulating
discussions.

\label{lastpage}
\end{document}